\documentclass[prd,a4paper,showpacs]{revtex4}
\usepackage{graphicx}
\usepackage{dcolumn}
\usepackage{amsmath}
\usepackage{bm}

\begin{document}

\preprint{Preprint Universit\'{e} de Mons-Hainaut}
\title{Semirelativistic potential model for glueball states}

\author{Fabian \surname{Brau}}
\thanks{FNRS Postdoctoral Researcher}
\email[E-mail: ]{fabian.brau@umh.ac.be}
\author{Claude \surname{Semay}}
\thanks{FNRS Research Associate}
\email[E-mail: ]{claude.semay@umh.ac.be}
\affiliation{Service de Physique G\'{e}n\'{e}rale et de Physique des
Particules \'{E}l\'{e}mentaires, Groupe de Physique Nucl\'{e}aire
Th\'{e}orique,
Universit\'{e} de Mons-Hainaut, Place du Parc 20,
B-7000 Mons, Belgium}

\date{\today}

\begin{abstract}
The masses of two-gluon glueballs are studied with a semirelativistic
potential model whose interaction is a scalar linear confinement
supplemented by a one-gluon exchange mechanism. The gluon is massless
but the leading corrections of the dominant part of the Hamiltonian are
expressed in terms of a state dependent constituent gluon mass. The
Hamiltonian depends only on 3 parameters: the strong coupling constant,
the string tension, and a gluon size which removes all singularities in
the leading corrections of the potential. Accurate numerical
calculations are performed with a Lagrange mesh method. The masses
predicted are in rather good agreement with lattice results and with
some experimental glueball candidates.
\end{abstract}


\maketitle

\section{Introduction}
\label{sec:intro}

The existence of bound states of gluons, called glueballs, is a
prediction of the QCD theory. The experimental discovery of such
particles would give a supplementary strong support to this theory. But,
a reliable experimental identification of glueballs is difficult to
obtain, mainly because glueball states might possibly mix strongly with
nearby
meson states. Nevertheless, the computation of pure gluon glueballs
remains an interesting task. This could guide experimental searches and
provide some calibration for more realistic models of glueballs.

The potential model, which is so successful to describe bound states of
quarks, is also a possible approach to study glueballs. Among the
pioneer works using this formalism, the one of Corwall and Soni is
particularly interesting \cite{corn83}. Assuming a nonrelativistic
kinematics and a saturated confinement supplemented by a one-gluon
exchange interaction, masses of pure gluon glueballs were computed.
However, only the four lightest two-gluon states ($0^{++}$, $0^{-+}$,
$1^{-+}$, $2^{++}$) were computed and found between 1.2 and 1.8~GeV.
Using a similar model, two-gluon glueballs have been studied in
Ref.~\cite{hou03}. The masses of states with $L=0$, 1 and 2 orbital
momentum were computed and several states are found below 3~GeV.
Unfortunately, we think that this model suffers from several drawbacks
which spoil any possible physical conclusions. This has been discussed
elsewhere \cite{brau04}. Nevertheless, we think that these drawbacks can
be corrected in order to obtain a more reliable potential model.

In this paper, we compute two-gluon glueball masses using various
modifications of the potential model obtained two decades ago by
Cornwall and Soni. After a critical study of these various models, we
conclude that a spectra in rather good
agreement with lattice calculations and some experimental glueball
candidates can be obtained, provided several conditions are fulfill: a
semirelativistic Hamiltonian is used, the gluon has a finite size, the
confinement is a scalar interaction, and a dynamical constituent gluon
mass is used in the leading relativistic corrections.

In Sec.~\ref{sec:ham}, the Hamiltonian model is presented with the
notion of gluon size. Three variants of our model are discussed in
Sec.~\ref{sec:res} and a glueball spectrum is presented. Concluding
remarks are given in Sec.~\ref{sec:conc} and some useful technical
details are given in appendix.

\section{Hamiltonian}
\label{sec:ham}

The two-gluon Hamiltonian contains a kinetic part $H_0$, a short-range
part $V_{\text{SR}}$ due to the one-gluon exchange between the two
valence gluons, and a confining interaction $V_{\text{conf}}$, as the
model proposed by Corwall and Soni \cite{corn83}
\begin{equation}
\label{h}
H = H_0 + V_{\text{SR}} + V_{\text{conf}}.
\end{equation}
Following Refs.~\cite{corn83,simo01}, there is no constant potential,
contrary to usual Hamiltonians for mesons and baryons.
This model can be considered with both nonrelativistic (Schr\"odinger
equation) and semirelativistic (spinless Salpeter equation) kinematics
($\hbar=c=1$)
\begin{equation}
\label{h0}
H_0 = 2 m_K +\frac{\bm p^2}{m_K}
\quad \text{or} \quad
H_0 = 2 \sqrt{\bm p^2 + m_K^2},
\end{equation}
where $m_K$ is the effective gluon mass appearing in the free part of
the Hamiltonian. If a Schr\"odinger Hamiltonian is used, it is necessary
to verify that, for each glueball state, the quantity
$\sqrt{\langle \bm p^2/ m_K^2 \rangle}$, which can be considered as the
mean speed of a gluon, is small with respect to 1.

\subsection{Short range potential}
\label{ssec:sr}

We use the short-range potential between two gluons proposed in
Ref.~\cite{corn83}. After some manipulations, this potential takes de
following form
\begin{eqnarray}
\label{vsr}
V_{\text{SR}} &=& -\lambda\, U(r) \left(
\frac{2 s - 7 m^2}{6 m^2} + \frac{1}{3} {\bm S}^2 \right) \nonumber \\
&&+ \frac{\lambda \pi}{3 m^2}\delta^3(\bm r) \left(
\frac{4 m^2 - 2 s}{m^2} + \frac{5}{2} {\bm S}^2 \right) \nonumber \\
&&- \frac{3 \lambda}{2 m^2} \frac{U'(r)}{r}\,\bm L \cdot \bm S
\nonumber \\
&&+ \frac{\lambda}{6 m^2} \left( \frac{U'(r)}{r} - U''(r) \right) T
\quad \text{with} \quad U(r)=\frac{e^{-m r}}{r},
\end{eqnarray}
where $\bm L$ and $\bm S$ are the usual orbital momentum and spin
operators, and where
\begin{equation}
\label{tens}
T = \bm S^2 - 3 (\bm S \cdot \hat{\bm  r})^2
\end{equation}
is the tensor operator. $m$ is
an effective gluon mass, which can differ from the mass $m_K$ (see
below). The quantity $s$ is in principle the square of the glueball
mass, but we will always take
$s=4m^2$ as it is suggested in Ref.~\cite{corn83}. The parameter
$\lambda$ is linked to the strong coupling
constant $\alpha_S$ by the relation \cite{hou84}
\begin{equation}
\label{lambda}
\lambda = 3 \alpha_S.
\end{equation}

This potential has a priori a very serious flaw: depending on the
spin state, the short distance singular parts of the potential may be
attractive and lead to a Hamiltonian unbounded from below. We will see
in Sec.~\ref{ssec:size}, how to cure this problem.

\subsection{Confinement potential}
\label{ssec:conf}

The dominant part of the interaction between the two gluons is the
confinement. As the leading relativistic corrections are taken into
account in the short-range part of the interaction, it is natural to
keep the same order corrections for the confinement potential. The
Lorentz structure of the confining interaction is not well known yet. In
this work, we follow the prescription of Ref.~\cite{simo00}: if the
radial form (static or zero order part) of the confinement is $W_X(r)$,
then the total confinement interaction is written
\begin{equation}
\label{vconf}
V_{\text{conf}}= W_X(r) - \frac{1}{2 m^2 r} W_X'(r) \,\bm L \cdot \bm S,
\end{equation}
where the effective mass $m$ is the same than the one appearing in
potential~(\ref{vsr}). Actually, this form contains only the dominant
correction, which is a spin-orbit contribution, and it is only valid for
large values of the distance $r$. But, for our purpose, this
approximation is sufficient. The interaction~(\ref{vconf}) corresponds
to a confinement with a dominant scalar structure \cite{luch91}. Let us
note that the spin-orbit contribution from the confinement counteracts
the spin-orbit contribution from the one-gluon exchange, and plays an
important role to obtain a spectra in agreement with lattice
calculations.

Two radial forms $W_X(r)$ can be used. In
Refs.~\cite{corn83,hou01,hou03}, the confinement potential saturates at
large distances
\begin{equation}
\label{confsat}
W_\beta(r) = 2m\left( 1-e^{-\beta m r} \right).
\end{equation}
Such a form can simulate the breaking of the color flux tube between the
two gluons due to color screening effects. The maximal mass for a
glueball is then $4 m$. Another simpler form is
proposed in Ref.~\cite{simo00}
\begin{equation}
\label{conflin}
W_a(r) = a_G\, r = \frac{9}{4} a\, r,
\end{equation}
where $a_G$ is the string tension between two gluons and $a$ the usual
string tension between a quark and an antiquark.
The $9/4$ factor is due to the color configuration of the
gluons \cite{corn83,hou03}. These two potentials coincide at small
distances, which implies that
\begin{equation}
\label{beta}
\beta \approx \frac{a_G}{2 m^2} = \frac{9 a}{8 m^2}.
\end{equation}
Potential~(\ref{conflin}) seems a priori inappropriate since strings
joining gluons must always break if a sufficiently high energy is
reached. But this phenomenon must only contribute to the masses of the
highest glueball states.

\subsection{Gluon size}
\label{ssec:size}

Within the framework of a potential QCD model, it is natural to assume
that a gluon is not a pure pointlike particle but an effective degree of
freedom which is dressed by a gluon and quark-antiquark pair cloud. Such
hypothesis for quarks leads to very good results in the meson
\cite{brau98} and baryon \cite{brau02} sectors. We assume here a Yukawa
color charge density for the gluon
\begin{equation}
\rho(\bm u)=\frac{1}{4\pi \gamma^2}
\frac{e^{-u/\gamma}}{u},
\label{dens}
\end{equation}
where $\gamma$ is the gluon size parameter.
The interaction between two gluons is then modified by this density, a
bare
potential $V$ being transformed into a dressed potential
$\widetilde{V}$.
This potential is obtained by a double convolution over the densities of
each interacting gluon and the potential. It can be shown that this
procedure is equivalent to the following calculation \cite{sema03}
\begin{equation}
\widetilde{V}(\bm r)=\int d\bm r'\,V(\bm r')\, \Gamma(\bm r-\bm r')
\quad \text{with} \quad
\Gamma(\bm u)=\frac{1}{8\pi \gamma^3}e^{-u/\gamma}.
\label{conv}
\end{equation}
Convolutions for some useful potentials are given in
appendix~\ref{sec:conv}.

Other color charge densities could be used, a Gaussian one for instance
\cite{brau02}. We have nevertheless strong indications that such a
change cannot noticeably modify the results \cite{brau01}. We choose
the Yukawa density because all convolutions are analytical with this
form.

The convolution~(\ref{conv}) of potential $e^{-m r}/r$ with the color
density~(\ref{dens}) removes all singularities in the short-range
interaction~(\ref{vsr}). For consistency, we apply the same
regularization to the confinement potential~(\ref{vconf}), although no
singularity are present in the radial form $W_X(r)$.

\section{Results}
\label{sec:res}

\subsection{Some general considerations}
\label{ssec:consid}

Nonrelativistic potential models have been intensively used to compute
static properties of mesons and baryons. But numerous works show that
semirelativistic potential models can give better results (see for
instance Refs.~\cite{sema92,fulc94}). Although the gluon effective mass
is expected around 700~MeV \cite{corn83,hou03}, which is heavier than
the assumed constituent strange quark mass, the relevance of a
nonrelativistic dynamics for the gluon is questionable. Using the
various models discussed below with several different sets of realistic
parameters, we have always obtained values of
$\sqrt{\langle \bm p^2/ m_K^2 \rangle}$ around unity (and sometimes
largely above) when a nonrelativistic kinematics was used (if $m_K\ne
0$). As we shall see below, the model~III is by nature a
semirelativistic one. Both kinematics can be used for models~I and II,
but we have verified that the drawbacks of these two models cannot be
solved by a change of kinematics. Consequently, in the following, we
will only present results from spinless Salpeter Hamiltonians.

In order to avoid singularities, potentials $U(r)$ and $W_X(r)$ of
formulas~(\ref{vsr}) and (\ref{vconf}) have been replaced by their
corresponding dressed forms $\widetilde{U}(r)$ and
$\widetilde{W}_X(r)$. Another way to get rid of singularities in the
short-range potential is
to treat $V_{\text{SR}}$ as a perturbation (at least when it is
attractive) of the dominant confinement interaction. But, for a lot of
states, computed with different sets of realistic parameters, the
contribution of the short-range part can be comparable to the one of the
confinement part. So, a perturbative approach of the singularities can
hardly be justified. In the following, we will always treat the
potential $V_{\text{SR}}$ non perturbatively.

The tensor operator $T$ is responsible of channel coupling. Its matrix
elements are given in appendix~\ref{sec:meop}. It can be shown that the
total spin $S$ of two gluons is always a good quantum numbers, but
mixing of orbital momenta with the same parity is possible. For
instance, the $2^{++}$ glueball with $S=2$ is the mixing of three states
with $L=0$, 2 and 4.  But, it is not coupled with the $2^{++}$ glueball
with $S=0$ and $L=2$. In principle, the effect of mixing is a second
order correction with respect to the contribution of the diagonal term.
In this paper, results are only shown when off-diagonal tensor
contributions are neglected. Nevertheless, we will discuss below the
effect of mixing for our various models.

The general characteristics of all our models are:
\begin{itemize}
\item Semirelativistic kinematics;
\item All radial forms convoluted following relation~(\ref{conv});
\item $V_{\text{SR}}$ not considered as a perturbation of
$V_{\text{conf}}$;
\item No channel coupling with $T$.
\end{itemize}

The eigenvalue problem has been solved by the Lagrange-mesh method which
allows a great accuracy as well as for Schr\"odinger equation
\cite{baye86} as for spinless Salpeter equation \cite{sema01}.

\begin{center}
\begin{figure}
\includegraphics*[height=8cm]{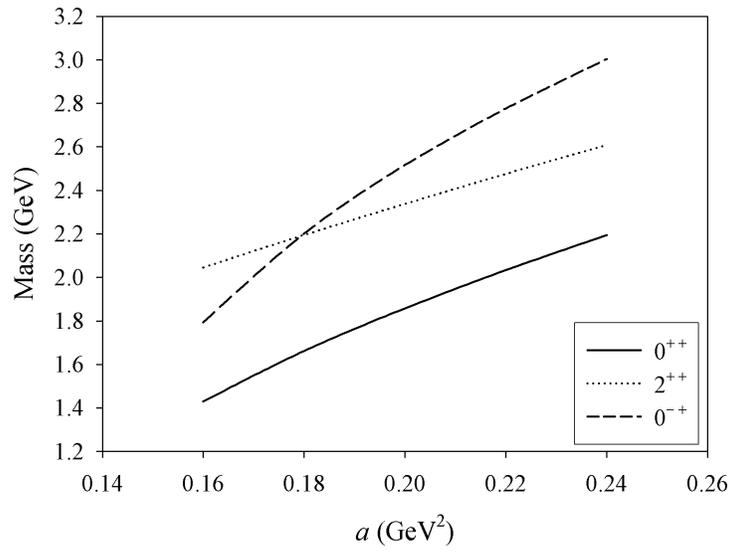}
\caption{Masses of glueball states $0^{++}$ ($L=S=0$), $2^{++}$
($L=0, S=2$), and $0^{-+}$ ($L=S=1$) as a function of the string tension
$a$, for the model III with $\alpha_S=0.5$ and $\gamma=0.52$
GeV$^{-1}$.}
\label{fig1}
\end{figure}
\end{center}

\begin{center}
\begin{figure}
\includegraphics*[height=8cm]{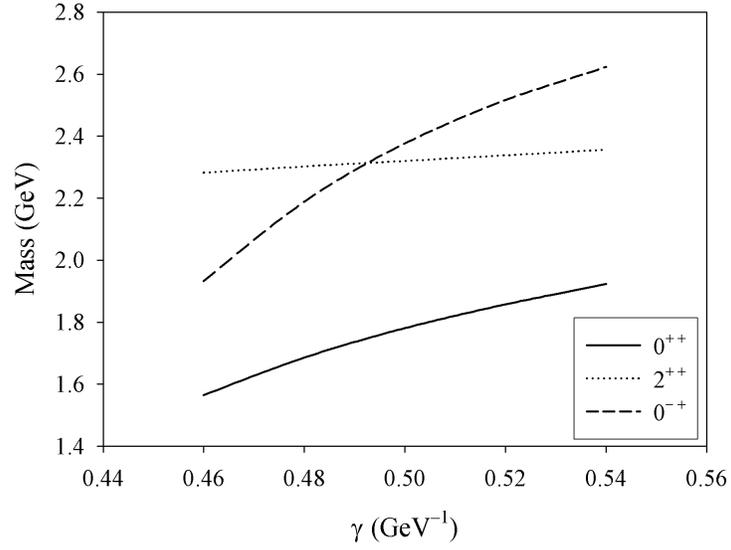}
\caption{Masses of glueball states $0^{++}$ ($L=S=0$), $2^{++}$
($L=0, S=2$), and $0^{-+}$ ($L=S=1$) as a function of the gluon size
$\gamma$, for the model III with $\alpha_S=0.5$ and $a=0.2$
GeV$^2$.}
\label{fig2}
\end{figure}
\end{center}

\begin{center}
\begin{figure}
\includegraphics*[height=8cm]{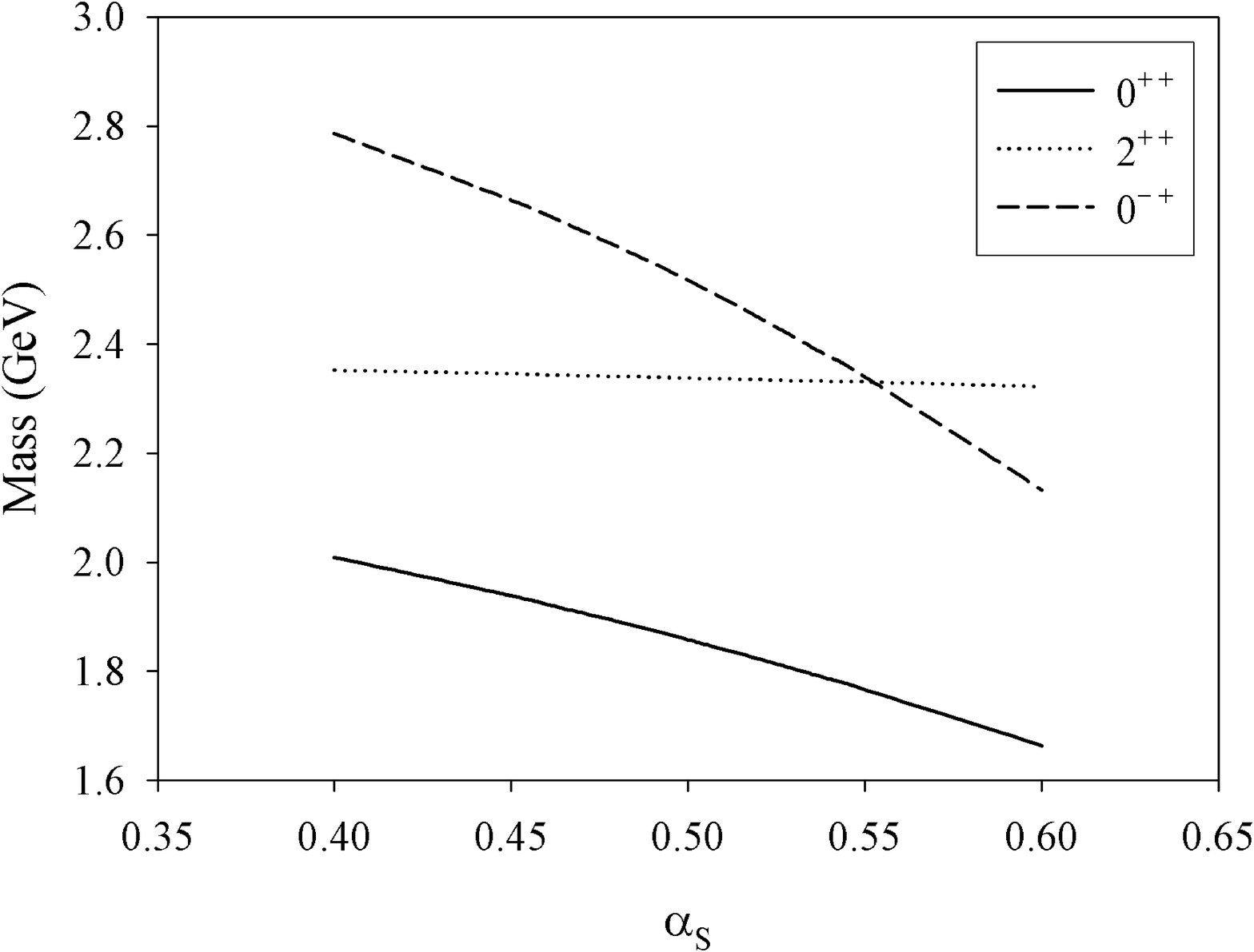}
\caption{Masses of glueball states $0^{++}$ ($L=S=0$), $2^{++}$
($L=0, S=2$), and $0^{-+}$ ($L=S=1$) as a function of the strong
coupling constant
$\alpha_S$, for the model III with $a=0.2$ GeV$^2$ and $\gamma=0.52$
GeV$^{-1}$.}
\label{fig3}
\end{figure}
\end{center}

In the models considered below, the gluon masses ($m_K$ and $m$) will be
fixed by physical considerations. We are then left with 3 parameters:
$\alpha_s$ (or $\lambda$) for $V_{\text{SR}}$, $\beta$ or $a$ for
$V_{\text{conf}}$, and the gluon size $\gamma$ for which less
constraints exist on its value. Fortunately, the mass of the lightest
$2^{++}$ state is nearly independent of the parameters $\alpha_s$ and
$\gamma$.
For this state, the spin-orbit and diagonal tensor potentials vanish and
the two remaining contributions have opposite signs. So, the confinement
potential is the largely dominant contribution to this $2^{++}$ mass
(see Figs.~\ref{fig1}-\ref{fig3}). We can fix the value of $\beta$ or
$a$ with this state only, knowing that a lattice calculation
\cite{morn99} and a quasiparticle model \cite{szcz03} favor a
$2^{++}$ mass around 2.4~GeV, but that some experimental
candidates are found around 2~GeV \cite{zou99,bugg00}. Some agreements
between theoretical calculations and experimental data exist about mass
ratios of some lightest glueball candidates (see
Refs.~\cite{morn99,zou99,bugg00} and Table~\ref{tab1}):
$M(0^{++})/M(2^{++}) \sim 0.72-0.78$ and
$M(0^{-+})/M(2^{++}) \sim 1.10$. So, we will fix the parameters
$\alpha_s$ and $\gamma$ in order to reproduce at best these mass ratios.
Let us note that the mass ratios in lattice calculations can be more
interesting quantities to consider than absolute masses, due to the
existence of normalization problems \cite{morn99}.

\subsection{Model I}
\label{ssec:m1}

The first model we consider is by two aspects close to the models of
Refs.~\cite{corn83} and \cite{hou03}: The short-range part is
supplemented by a saturated confinement potential and the gluon is
assumed to be characterized by an unique effective mass $m_K=m$, around
the typical value of 700~MeV. In this paper, we choose the value of
Ref.~\cite{hou03}. Nevertheless, the model is
semirelativistic and the short-range part is not treated perturbatively.
The particular characteristics of model I are:
\begin{itemize}
\item $m_K = m = 0.670$ GeV \cite{hou03};
\item $V_{\text{conf}}$ with $W_\beta(r)$.
\end{itemize}

To find a mass of the lightest $2^{++}$ state around 2~GeV, it is
necessary to take $\beta \approx 0.2$, which corresponds to
$a\approx 0.08$~GeV$^2$, a quite unrealistic value. A mass around
2.4~GeV can be obtained with $\beta\approx 0.5$ corresponding to
$a\approx 0.2$~GeV$^2$, a more realistic value. The parameters
$\beta=0.5$, $\alpha_S = 0.5$, and $\gamma = 0.5$~GeV$^{-1}$
give the following lightest mass ratios: $M(0^{++})/M(2^{++}) = 0.93$
and $M(0^{-+})/M(2^{++}) = 0.73$. Let us note that if the spin-orbit
contribution from confinement is not taken into account, the last ratio
decreases to 0.61. A lower value for the mass ratio
$M(0^{++})/M(2^{++})$ can be obtained by modifying the parameters
$\alpha_S$ and $\gamma$, but in this case, the mass ratio
$M(0^{-+})/M(2^{++})$ also decreases. It can even take negative values
for realistic values of the parameters $\alpha_S$ and $\gamma$. This
nonphysical behavior is due to the spin-orbit potential from
$V_{\text{SR}}$ which becomes very attractive. It is worth noting that a
variation of the gluon effective mass of 200~MeV around the value
670~MeV does not change noticeably the results. Thus, the model~I is not
able to describe neither the lightest $2^{++}$ mass state, nor the
lightest
mass ratios $M(0^{++})/M(2^{++})$ and $M(0^{-+})/M(2^{++})$.

If the channel mixing due the tensor operator is turned on, the
situation gets worse. For instance, without channel mixing the
lightest $2^{++}$ state ($L=0$, $S=2$) has a reasonable mass. With
channel mixing, the $L=0$ component is coupled with $L=2$ and $L=4$
components for which the total spin-orbit contribution is very
attractive. The mass of this state becomes then negative, even for
realistic values of the parameters $\beta$, $\alpha_S$ and $\gamma$.

\subsection{Model II}
\label{ssec:m2}

The model II is the same as model I but with the saturated confinement
replaced by the linear confinement. The particular characteristics of
model II are:
\begin{itemize}
\item $m_K = m = 0.670$ GeV \cite{hou03};
\item $V_{\text{conf}}$ with $W_a(r)$.
\end{itemize}
This model and the previous one give essentially the same results about
the lightest mass ratios $M(0^{++})/M(2^{++})$ and
$M(0^{-+})/M(2^{++})$. Moreover,
masses around 2.4~GeV and 2~GeV are obtained for the
lightest $2^{++}$ state with $a \approx 0.12$~GeV$^2$ and
$a \approx 0.07$~GeV$^2$ respectively, quite unrealistic values of the
meson string tension.

\subsection{Model III}
\label{ssec:m3}

With the two previous models, the spin-orbit effect from the one
gluon-exchange is too attractive and cannot be
counteracted efficiently by the spin-orbit contribution coming from the
confinement. The strength of this attractive potential can be reduced by
decreasing the values of $\alpha_S$. But in this case, it is not
possible to obtain reasonable mass ratios for the lightest $0^{++}$,
$2^{++}$, and $0^{-+}$ states. Another possibility is to increase the
values of the effective mass $m$. But, if we keep the link $m_K=m$, then
too high masses are obtained for all glueballs, due to the contribution
of the kinetic part of the Hamiltonian. Fortunately, it is
physically relevant to choose $m_K \ll m$.

For a system of two identical particles with mass $m$, the coefficient
$1/m^2$ appears naturally in the relativistic corrections of a static
potential. A better approximation, proposed in Ref.~\cite{luch91} and
used for instance in Ref.~\cite{godf85}, is to replace this
coefficient by $1/E^2(\bm{p})$ where $E(\bm{p})=\sqrt{\bm{p}^2+m^2}$.

A similar procedure is also proposed within the auxiliary field
formalism (also called einbein field formalism) \cite{morg99}, which can
be considered as an approximate way to handle semirelativistic
Hamiltonians \cite{sema04}. Within this approach, the
effective QCD Hamiltonian for two identical particles (quark or gluon)
depends on the current particle mass $m$ and on a state dependent
constituent mass $\mu = \sqrt{\langle \bm{p}^2+m^2 \rangle}$
\cite{sema04}. All corrections to the static potential are then expanded
in powers of $1/\mu^2$.

We will adapt these prescriptions in the model III. In principle, the
mass appearing in the leading corrections must be replaced by the
operator $E(\bm{p})=\sqrt{\bm{p}^2+m^2}$ where $m$ is the mass appearing
in the kinetic operator. This leads to a very complicated nonlocal
potential which is very difficult to handle. So we will use an
approximation.

Following the hypotheses of Ref.~\cite{simo00}, we will assume that the
gluons are massless and that the dominant effective QCD Hamiltonian
$H_{gg}$ for a two-gluon glueball is written
\begin{equation}
\label{hgg}
H_{gg}=2 \sqrt{\bm p^2} + a_G\, r.
\end{equation}
The eigenvalues $M_{nL}$ and the constituent masses $\mu_{nL}$
corresponding to this Hamiltonian are in very good approximation given
by the following relations \cite{simo89,sema04}
\begin{equation}
\label{mu}
M_{nL} = 4 \mu_{nL} \quad \text{with} \quad
\mu_{nL} = \sqrt{a_G} \left( \frac{\epsilon_{nL}}{3} \right)^{3/4}.
\end{equation}
$\epsilon_{nL}$ is an eigenvalue of the Hamiltonian
$\bm q^2 + |\bm x|$, in which $\bm q$ and $\bm x$ are dimensionless
conjugate variables. These eigenvalues can be
accurately computed with a Lagrange-mesh method for instance
\cite{baye86}. Let us note that $\epsilon_{n0}$ is the $n$th zero of the
Airy function. These constituents masses will then appear in the leading
corrections to the dominant hamiltonian $H_{gg}$.

Finally, the particular characteristics of model III are:
\begin{itemize}
\item $m_K = 0$ and $m = \mu_{nL}$ (\ref{mu});
\item $V_{\text{conf}}$ with $W_a(r)$.
\end{itemize}
Let us note that, with our hypotheses, the constituent mass depends on
the principal quantum number $n$ of the glueball, on its orbital
momentum $L$, but not on the quantum numbers $S$ and $J$.

As mentioned before, we fix the value of the string tension only with
the mass of the lightest $2^{++}$ glueball. Equations~(\ref{mu}) show
that, in first approximation, the mass scale is simply given by
$\sqrt{a_G}$. By computing a great number
of spectra for various parameters, we have remarked that the values of
the lightest mass ratios $M(0^{++})/M(2^{++})$ and $M(0^{-+})/M(2^{++})$
cannot be fixed independently. Provided an approximate linear dependence
is kept between the two parameters $\alpha$ and $\gamma$, these mass
ratios do not change significantly. Finally, we have chosen to present
the glueball spectra for two sets of parameters.

\begin{center}
\begin{figure}
\includegraphics*[height=8cm]{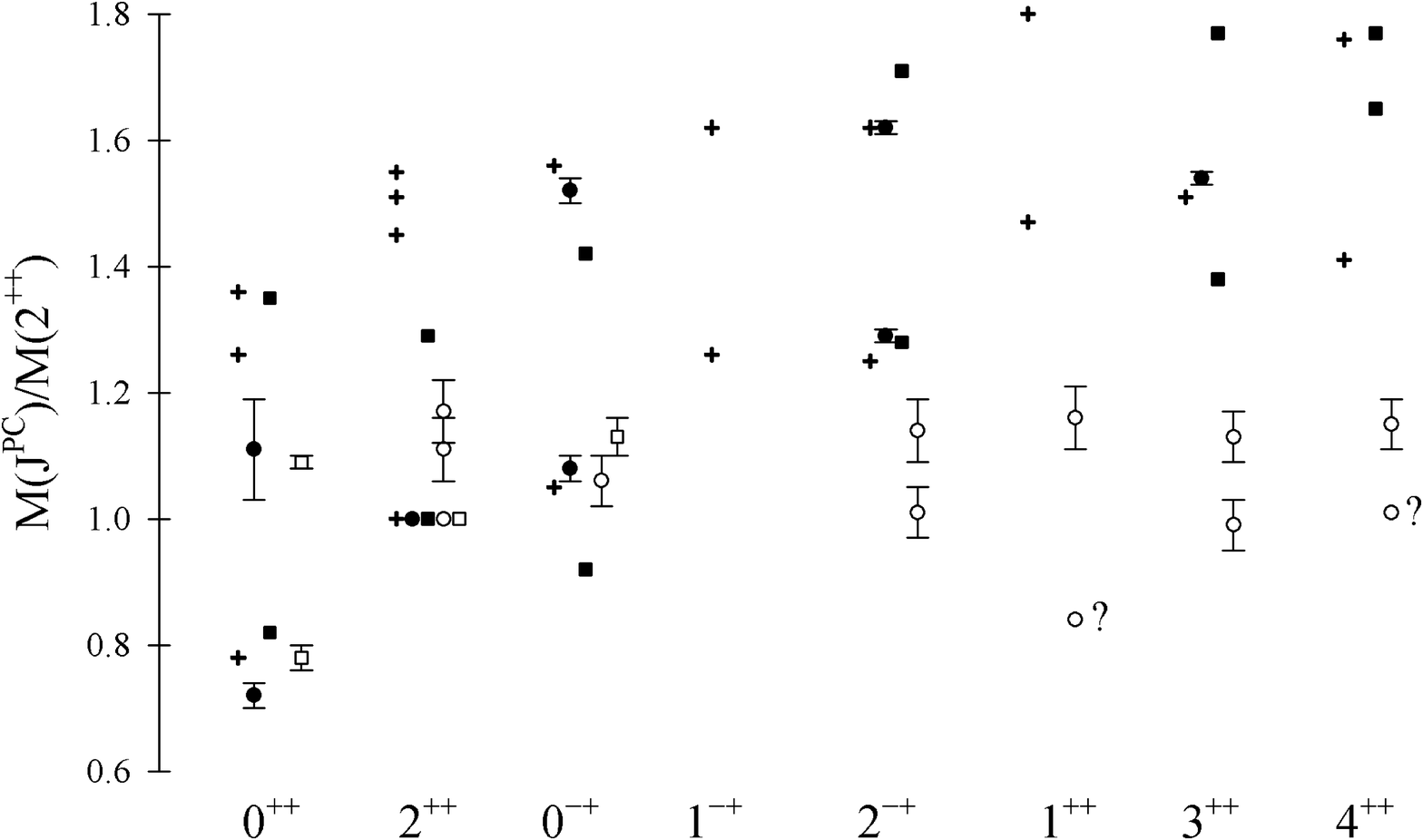}
\caption{Glueball mass ratios normalized to the lightest $2^{++}$ state
(see Table~\ref{tab1}).
Cross: Set of parameters B for model III;
Black circle: Lattice results \cite{morn99};
Black square: Quasiparticle model \cite{szcz03};
White circle: Experiment \cite{zou99};
White square: Experiment \cite{bugg00}.
The experimental states with a question mark are seen but the
uncertainty is not known.
}
\label{fig4}
\end{figure}
\end{center}

With $a=0.16$~GeV$^2$, $\alpha_S = 0.40$, and $\gamma=0.504$~GeV$^{-2}$
(set A),
the mass of the lightest $2^{++}$ glueball is 2051~MeV, which is
close to some experimental candidates. These
values for $a$ and $\alpha_S$ are near those used in some recent baryon
calculations \cite{naro02}. Moreover, the value of the string tension is
close to a value found in a recent lattice study \cite{taka01}. With
$a=0.21$~GeV$^2$, $\alpha_S=0.50$, and $\gamma=0.495$~GeV$^{-1}$ (set
B), the mass of the lightest $2^{++}$ glueball is 2384~MeV, which is
close to a result obtained with lattice calculations. All results
are presented in table~\ref{tab1} and in Fig.~\ref{fig4} with some
results from a lattice
calculations \cite{morn99} and from a quasiparticle approach with no
free parameters \cite{szcz03}, and with some possible experimental
candidates \cite{zou99,bugg00}.

We can see that the mass ratios for the two sets of parameters are in
rather good agreement with the theoretical mass ratios predicted by the
lattice study \cite{morn99}. Moreover, the absolute masses for set B are
within the theoretical error bars of the lattice masses. The largest
discrepancy is for the first excited $0^{++}$ glueball, the state
predicted by the lattice calculations with the largest error bar.

Our mass ratios are similar to those obtained in the quasiparticle model
\cite{szcz03}, but are closer to the mass ratios of the lattice studies.
It is worth mentioning that this quasiparticle approach contains no free
parameters. Again, it favors a 2.4~GeV value for the mass of the
lightest $2^{++}$ state, like the lattice model.

The lack of reliable identification of glueballs makes comparison with
experimental data more hazardous. The set A results are in rather good
agreement with data only for the lightest $0^{++}$, $2^{++}$, and
$0^{-+}$ glueball candidates, the states used to fix the parameters. The
general tendance of our model and of lattice calculations is to predict
excited states with higher masses than those which seem to be observed.

Lattice calculations seem to rule out the presence of $1^{-+}$ and
$1^{++}$ states below 4~GeV. This can be qualitatively understood in
terms of interpolating operators of minimal dimension which can create
glueball states, with the expectation that higher dimensional operators
create higher mass states: the lowest states $0^{++}$, $2^{++}$,
$0^{-+}$ and $2^{-+}$ are produced by dimension-4 operators, while
$1^{++}$ and $1^{-+}$ are respectively produced by dimension-5 and
dimension-6 operators \cite{morn99}. Nevertheless, our model predicts
the existence of $1^{-+}$ and $1^{++}$ states around 3~GeV. In the model
of Ref.~\cite{corn83}, a $1^{-+}$ state is predicted below the $2^{++}$
state (no $1^{++}$ state is mentioned). Possible experimental candidates
exist for low mass $1^{++}$ states \cite{zou99}, but as mentioned above,
the identification is far from certain. The presence of the relatively
low mass $1^{-+}$ and $1^{++}$ states in our model may be due to the use
of massive valence gluons with three states of polarization (creation of
a spin one glueball with two massless gluons, with only two states of
polarization, is problematical). In our model III, the gluon is
massless in the kinetic part, but a constituent non-zero mass
unavoidably appears for the spin corrections \cite{simo00}. The presence
of spin one states around 3~GeV in our model could indicate what are the
limits of a potential approach.

If the spin-orbit contribution from confinement is not taken into
account, the agreement between our masses and the lattice results become
poorer. For the parameters of set A, the mass ratio for the lightest
$0^{-+}$ glueball changes from 1.06 to 0.87, and the mass ratio for the
first excited $0^{++}$ glueball changes from 1.26 to 0.83. This shows
that the spin-orbit contribution from confinement is an important
ingredient of the model.

The channel mixing due to the tensor force is difficult to implement
within this model. As $\mu$ depends on the orbital momentum, the
diagonal potential for each channel is characterized by a different
value of $\mu$. The problem is to define this parameter for the mixing
potentials. We have performed several test computations using a mean
value of $\mu$ for all channels. This gave us strong indications that
the coupling of channels has a small influence on the glueball masses,
contrary to the two previous models. We estimate that the masses of the
lightest glueballs could be modified by a quantity comprised between 50
and 100~MeV.

\section{Conclusion}
\label{sec:conc}

The masses of pure two-gluon glueballs have been studied with a
semirelativistic potential model. The potential is the sum of a
one-gluon exchange interaction and a linear confining potential, assumed
to be of scalar type. The gluon is massless but the leading corrections
of the dominant part of the Hamiltonian are expressed in terms of a
state dependent constituent mass. The Hamiltonian depends only on 3
parameters: the strong coupling constant, the string tension, and the
gluon size. This last parameter, less constrained than the two others by
the QCD theory, removes all singularities in the leading corrections of
the potential. These corrections are not treated as perturbations of
the dominant part. All masses have been accurately computed with a
Lagrange mesh method.

The masses predicted by our potential model are in agreement with
experimental glueball candidates only for the lightest $0^{++}$,
$2^{++}$, and $0^{-+}$ states \cite{zou99,bugg00}, but are in rather
good agreement with spectra obtained by a lattice calculation
\cite{morn99} and in reasonable agreement with spectra
obtained by a quasiparticle model \cite{szcz03}. A notable
difference is the presence in our model of spin one states around 3~GeV.
This could indicate the limit of the validity for a potential approach.

We have tested other nonrelativistic and semirelativistic potential
models in which a constant constituent gluon mass is used, and we have
found that it is not possible to obtain good spectra for realistic
values of the QCD parameters (see
Sec.~\ref{ssec:m1} and \ref{ssec:m2}). The main problem
arises from the strongly attractive spin-orbit potential for the
one-gluon exchange. When its strength is not reduced by a large
constituent gluon mass, it can lead to negative nonphysical glueball
mass.

The constituent gluon mass is introduced in our model by an approximate
procedure which relies on the existence of a pure linear confinement
between the gluons \cite{simo89,sema04}. A more physical ansatz should
be to define the constituent mass as a momentum dependent operator
($\sqrt{\bm{p}^2}$ for massless gluon) \cite{luch91}. It could then be
possible to take into account correctly the channel coupling due to the
tensor forces, and to use naturally a saturated confinement potential.
It could also be interesting to compute three-gluon glueball masses
within the same model. Such a work is in progress.

\section{Acknowledgments}

F. Brau (FNRS Postdoctoral Researcher) and C. Semay (FNRS Research
Associate) would like to thank the FNRS for financial support.

\appendix

\section{Convolutions}
\label{sec:conv}

Applied for some useful potentials, the formula~(\ref{conv}) gives
\begin{eqnarray}
\label{convdel}
\widetilde{\delta^3}(\bm r) &=& \frac{1}{8\pi \gamma^3}e^{-r/\gamma}, \\
\label{convlin}
\widetilde{r} &=& r + \frac{4\gamma^2}{r} \left( 1-e^{-r/\gamma} \right)
-\gamma e^{-r/\gamma}, \\
\label{convgauss}
\widetilde{e^{-a r}} &=& \frac{4 a \gamma^2}{(a^2\gamma^2-1)^3}
\left( \frac{e^{-a r}}{r} - \frac{e^{-r/\gamma}}{r} \right) +
\frac{e^{-a r} + a \gamma \,e^{-r/\gamma}}{(a^2\gamma^2-1)^2},\\
\label{convyuk}
\widetilde{\frac{e^{-a r}}{r}} &=& \frac{1}{(a^2\gamma^2-1)^2}
\left( \frac{e^{-a r}}{r} - \frac{e^{-r/\gamma}}{r} \right) +
\frac{e^{-r/\gamma}}{2\gamma(a^2\gamma^2-1)}.
\end{eqnarray}
One can easily verify that
$\lim_{\gamma\rightarrow 0} \widetilde{U(r)} = U(r)$
for each potential $U(r)$.

\section{Angular momentum operators}
\label{sec:meop}

A system of two particles, with spin $s_1$ and $s_2$ respectively, with
a total spin $S$ and a total orbital angular momentum $L$  coupled to a
total angular momentum $J$, is noted here
$|L, S \rangle = |s_1, s_2; L, S; J \rangle$.
The mean value of the operators $\bm S^2$ and $\bm L \cdot \bm S$ are
trivial to compute
\begin{eqnarray}
\label{ops2l2}
\langle L', S'| \bm S^2  |L, S \rangle &=&
S(S+1) \,\delta_{L',L}\, \delta_{S',S},  \\
\langle L', S'| \bm L \cdot \bm S  |L, S \rangle &=& \frac{1}{2}
\Big[ J(J+1) - L(L+1) - S(S+1) \Big] \,\delta_{L,L'}\, \delta_{S,S'}.
\end{eqnarray}
The computation of the mean value of the operator $T$ is much more
involved. Using formulas from Ref.~\cite{vars88}, one can find
($\hat n = \sqrt{2 n+1}$)
\begin{eqnarray}
\label{opt}
\langle L', S'| T |L, S \rangle &=& (-1)^{L+L'+S'+J}\, \hat S \hat S'
\hat L \hat L'
\left( \begin{array}{ccc}
L & 2 & L' \\ 0 & 0 & 0 \end{array} \right)
\left \{ \begin{array}{ccc}
S & L & J \\ L' & S' & 2 \end{array} \right \} \\
&& \times \Bigg[ \phantom{+} (-1)^{S+1+s_2-3 s_1} \, \hat s_1
\sqrt{s_1 (s_1+1)  (2 s_1-1) (2 s_1+3)}
\left \{ \begin{array}{ccc}
s_1 & s_2 & S \\ S' & 2 & s_1 \end{array} \right \} \nonumber \\
&& \phantom{\times} \phantom{\Bigg[} + (-1)^{S+1+s_1-3 s_2} \, \hat s_2
\sqrt{s_2 (s_2+1)  (2 s_2-1) (2 s_2+3)}
\left \{ \begin{array}{ccc}
s_2 & s_1 & S \\ S' & 2 & s_2 \end{array} \right \} \nonumber \\
&& \phantom{\times} \phantom{\Bigg[} - 2\sqrt{30} \, \hat s_1 \hat s_2
\sqrt{s_1 (s_1+1) s_2 (s_2+1)} \left \{ \begin{array}{ccc}
s_1 & s_2 & S \\ 1 & 1 & 2 \\ s_1 & s_2 & S' \end{array} \right \}
\Bigg] . \nonumber
\end{eqnarray}


\begin{table}[h]
\protect\caption{Glueball masses in MeV (mass ratios normalized to
lightest $2^{++}$) [$L$ and $S$ quantum numbers only relevant for our
model] obtained with model III for two sets of parameters (A:
$a=0.16$~GeV$^2$, $\alpha_S=0.40$, and $\gamma=0.504$~GeV$^{-1}$; B:
$a=0.21$~GeV$^2$, $\alpha_S=0.50$, and $\gamma=0.495$~GeV$^{-1}$). Some
theoretical results from other models and some possible experimental
candidates are also indicated. The lightest $0^{++}$, $2^{++}$, and
$0^{-+}$ states are taken as inputs to fix the parameters.}
\label{tab1}
\begin{ruledtabular}
\begin{tabular}{llllllll}
\multicolumn{1}{c}{$J^{PC}$} & \multicolumn{3}{c}{Model III}  &
\multicolumn{1}{c}{Lattice \cite{morn99}} &
\multicolumn{1}{c}{Quasiparticle \cite{szcz03}} &
\multicolumn{1}{c}{Experiment \cite{zou99}} &
\multicolumn{1}{c}{Experiment \cite{bugg00}} \\
& \multicolumn{1}{c}{[$L$,$S$]} &
\multicolumn{1}{c}{A} & \multicolumn{1}{c}{B} & & & \\
\hline
$0^{++}$ & [0,0] & 1604 (0.78) & 1855 (0.78) &
1730$\pm$50$\pm$80 (0.72) & 1980 (0.82) &  & 1507$\pm$5 (0.78) \\

& [2,2]          & 2592 (1.26) & 2992 (1.26) &
2670$\pm$180$\pm$130 (1.11) & 3260 (1.35) & & 2105$\pm$15 (1.09) \\

& [0,0]          & 2814 (1.37) & 3251 (1.36) &
& & & \\

$2^{++}$ & [0,2] & 2051 (1.00) & 2384 (1.00) &
2400$\pm$25$\pm$120 (1.00) & 2420 (1.00) & 2020$\pm$50 (1.00) &
1934$\pm$12 (1.00) \\

& [0,2]          & 2985 (1.46) & 3447 (1.45) &
& 3110 (1.29) & 2240$\pm$40 (1.11) & \\

& [2,0]          & 3131 (1.53) & 3611 (1.51) &
& & 2370$\pm$50 (1.17) & \\

& [2,2]          & 3230 (1.57) & 3695 (1.55) &
& & & \\

$0^{-+}$ & [1,1] & 2172 (1.06) & 2492 (1.05) &
2590$\pm$40$\pm$130 (1.08) & 2220 (0.92) & 2140$\pm$30 (1.06) &
2190$\pm$50 (1.13) \\

& [1,1]          & 3228 (1.57) & 3714 (1.56) &
3640$\pm$60$\pm$180 (1.52) & 3430 (1.42) & & \\

$1^{-+}$ & [1,1] & 2626 (1.28) & 3011 (1.26) &
& & & \\

& [1,1]          & 3349 (1.63) & 3852 (1.62) &
& & & \\

$2^{-+}$ & [1,1] & 2573 (1.25) & 2984 (1.25) &
3100$\pm$30$\pm$150 (1.29) & 3090 (1.28) & 2040$\pm$40 (1.01) & \\

& [1,1]          & 3345 (1.63) & 3862 (1.62) &
3890$\pm$40$\pm$190 (1.62) & 4130 (1.71) & 2300$\pm$40 (1.14) & \\

$1^{++}$ & [2,2] & 3098 (1.51) & 3501 (1.47) &
& & $\sim$~1700 (0.84) & \\

& [2,2]          & 3753 (1.83) & 4294 (1.80) &
& & 2340$\pm$40 (1.16) & \\

$3^{++}$ & [2,2] & 3132 (1.53) & 3611 (1.51) &
3690$\pm$40$\pm$180 (1.54) & 3330 (1.38) & 2000$\pm$40 (0.99) & \\

& [2,2]          & 3762 (1.83) & 4332 (1.82) &
& 4290 (1.77) & 2280$\pm$30 (1.13) & \\

$4^{++}$ & [2,2] & 2897 (1.41) & 3360 (1.41) &
& 3990 (1.65) & 2044$\pm$? (1.01)  & \\

& [2,2]          & 3633 (1.77) & 4197 (1.76) &
& 4280 (1.77) & 2320$\pm$30 (1.15) & \\
\end{tabular}
\end{ruledtabular}
\end{table}


\begin{thebibliography}{aa}

\bibitem{corn83} J. M. Cornwall and A. Soni, Phys. Lett. B {\bf 120},
431 (1983).
\bibitem{hou03} W. S. Hou and G. G. Wong, Phys. Rev. D {\bf 67}, 034003
(2003).
\bibitem{brau04} F. Brau and C. Semay, \emph{Comment on ``Glueball
spectrum from a potential model"}, submitted to Phys. Rev. D.
\bibitem{simo01} Yu. A. Simonov, Phys. Lett. {\bf 515}, 137 (2001).
\bibitem{hou84} W. S. Hou and A. Soni, Phys. Rev. D {\bf 29}, 101
(1984).
\bibitem{simo00} Yu. A. Simonov,
in
\emph{Proceedings of the XVII Autumn School
Lisboa, Portugal, 24 September - 4 October 1999}, edited by L. Ferreira,
P. Nogueira, and J. I. Silva-Marco (World Scientific,
Singapore, 2000), p.~60; hep-ph/9911237.
\bibitem{luch91} W. Lucha, F. F. Sch\"oberl, and D. Gromes, Phys. Rep.
{\bf 200}, 127 (1991).
\bibitem{hou01} W. S. Hou, C. S. Luo, and G. G. Wong, Phys. Rev. D
{\bf 64}, 014028 (2001).
\bibitem{brau98} F. Brau and C. Semay, Phys. Rev. D {\bf 58}, 034015
(1998).
\bibitem{brau02} F. Brau, C. Semay, and B. Silvestre-Brac,
Phys. Rev. C {\bf 66}, 055202 (2002).
\bibitem{sema03} B. Silvestre-Brac, F. Brau, and C. Semay,
J. Phys. G: Nucl. Part. Phys. {\bf 29}, 2685 (2003).
\bibitem{brau01} F. Brau, Thesis, Universit\'{e} de Mons-Hainaut, 2001
(unpublished).
\bibitem{sema92} C. Semay and B. Silvestre-Brac,
Phys. Rev. D {\bf 46}, 5177 (1992).
\bibitem{fulc94} L. P. Fulcher, Phys. Rev. D {\bf 50}, 447 (1994).
\bibitem{baye86} D. Baye and P.-H. Heenen, J. Phys. A {\bf 19},
2041 (1986); D. Baye, J. Phys. B {\bf 28}, 4399 (1995).
\bibitem{sema01} C. Semay, D. Baye, M. Hesse, and B. Silvestre-Brac,
Phys. Rev. E {\bf 64}, 016703 (2001).
\bibitem{morn99} C. J. Morningstar and M. J. Peardon, Phys. Rev. D
{\bf 60}, 034509 (1999).
\bibitem{szcz03} A. P. Szczepaniak and E. S. Swanson, Phys. Lett. B
{\bf 577}, 61 (2003).
\bibitem{zou99} B. S. Zou, Nucl. Phys. {\bf A655}, 41 (1999).
\bibitem{bugg00} D. V. Bugg, M. J. Peardon, and B. S. Zou, Phys. Lett. B
{\bf 486}, 49 (2000).
\bibitem{godf85} S. Godfrey and N. Isgur, Phys. Rev. D {\bf 32}, 189
(1985).
\bibitem{morg99} V. L. Morgunov, A. V. Nefediev, and Yu. A. Simonov,
Phys. Lett. B {\bf 459}, 653 (1999).
\bibitem{sema04} C. Semay, B. Silvestre-Brac, and I. M. Narodetskii,
Phys. Rev. D {\bf 69}, 014003 (2004).
\bibitem{simo89} Yu. A. Simonov, Phys. Lett. B {\bf 226}, 151 (1989).
\bibitem{naro02} I. M. Narodetskii and M. A. Trusov, Yad. Fiz.
{\bf 65}, 949 (2002); Phys. Atom. Nucl. {\bf 65}, 917 (2002);
Phys. Atom. Nucl. {\bf 66} (2003), in print, hep-ph/0307131
\bibitem{taka01} T. T. Takahashi, H. Matsufuru, Y. Nemoto, and H.
Suganuma, Phys. Rev. Lett. {\bf 86}, 18 (2001); Phys. Rev. D {\bf 65},
114509 (2002).
\bibitem{vars88} D. A. Varshalovich, A. N. Moskalev, and V. K.
Khersonskii, \emph{Quantum Theory of Angular Momentum} (World
Scientific, Singapore, 1988).

\end{thebibliography}
\end{document}